# Nanosize effect: Enhanced compensation temperature and existence of magneto-dielectric coupling in SmFeO$_3$

Smita Chaturvedi,[1,4*] Priyank Shyam,[2] Rabindranath Bag,[1] Mandar M. Shirolkar,[3] Jitender Kumar,[1] Harleen Kaur,[1] Surjeet Singh,[1,4] A.M. Awasthi,[5] and Sulabha Kulkarni[6*]

[1]*Indian Institute of Science Education and Research, Pune, Dr. Homi Bhabha Road, Pashan, Pune- 411008, India*
[2]*Interdisciplinary Nanoscience Center, Aarhus University, Gustav Wieds Vej 14, Aarhus, Denmark*
[3]*Hefei National Laboratory for Physical Sciences at the Microscale, University of Science and Technology of China, Hefei, Anhui- 230026, People's Republic of China*
[4]*Centre for Energy Science, Indian Institute of Science Education and Research, Pune- 411008, India*
[5]*UGC-DAE Consortium for Scientific Research, University Campus, Khandwa Road, Indore- 452001, India*
[6]*Centre for Materials for Electronics Technology, Panchawati Road, Pune- 411008, India*

*Corresponding authors:smita.chaturvedi24@gmail.com,smita.chaturvedi@iiserpune.ac.in, s.kulkarni@iiserpune.ac.in*

## ABSTRACT

In transition metal oxides, quantum confinement arising from a large surface to volume ratio often gives rise to novel physico-chemical properties at nanoscale. Their size dependent properties have potential applications in diverse areas, including therapeutics, imaging, electronic devices, communication systems, sensors, and catalysis. We have analyzed structural, magnetic, dielectric, and thermal properties of weakly ferromagnetic SmFeO$_3$ nanoparticles of sizes about 55 nm and 500 nm. The nano-size particles exhibit several distinct features that are neither observed in their larger-size variants nor reported previously for the single crystals. In particular, for the 55 nm particle, we observe six-fold enhancement of compensation temperature, an unusual rise in susceptibility in the temperature range 550 to 630 K due to spin pinning, and coupled antiferromagnetic-ferroelectric transition, directly observed in the dielectric constant.

## I. INTRODUCTION

Nanoscale effects on the structure and magnetic/ferroelectric properties of multiferroic and ferroelectric transition metal oxides have been the focus of many recent investigations in materials science. This interest is due to the potential applications of such nanostructured transition metal oxides in various nanoscale devices. The micro/nano-structure of a material plays a significant role in influencing the physical properties. The effect of size and morphology on the optical, catalytic, magnetic and electric behaviors for various transition metal oxides have become active areas of investigation. In this context investigating, understanding, and controlling various aspects of particle size on the magnetic/ferroelectric properties in transition metal oxide nanoparticles (NPs) is a challenging and intriguing area of research. In NPs, the predominant contribution to the magnetization is from the surface spins because of their lower coordination.[1] Considerable variation of the magnetic and dielectric properties with change in the particle size is expected because of the associated changes in the relative number of surface spins.

Among the various transition metal oxides, rare-earth orthoferrite $RFeO_3$ (R is rare-earth element) compounds show extraordinary physical properties of spin-switching and magnetization-reversal, tunable by the applied magnetic field and/or temperature.[2] These ferrites also exhibit magnetic,[3] magneto- optic,[4] and multiferroic[5] properties. $RFeO_3$ compounds, particularly those which contain two types of magnetic ions, exhibit complex magnetic behaviour as a function of temperature, pressure, particle size, and magnetic field.[6,7] The exchange interaction between the transition metal ions is typically strong and antiferromagnetic in nature; as a result, the magnetic ordering in the transition metal sublattice typically takes place at higher temperatures than in the rare earth sublattice. Their magnetic behaviour is also less anisotropic compared to that of the rare earth sublattice, due to the quenching of the orbital angular momentum. This gives rise to an interesting scenario, where the highly anisotropic rare earth spin controls the orientation of the transition metal spins, resulting in complex magnetic structures.[8] $SmFeO_3$ (SFO) represents such a system. It has an orthorhombic crystal structure [Pnma/Pbnm ($D^{16}_{2h}$) space group] and is composed of four distorted perovskite unit cells. SFO shows a high magneto-striction coefficient, high magnetic ordering temperature ~ 670 K, and a high spin-reorientation temperature, which makes it a potential candidate for magneto-electric applications.[9] However, the possibility of improper ferroelectricity in SFO single crystals has been debatable.[10,11]

SFO is reported to show a significant change of magnetic behaviour at lower temperature, where the net magnetization of Sm- sublattice completely compensates the net magnetization due to the Fe-sublattice. Magnetic compensation has recently gained interest due to its possibilities for both information storage and thermomagnetic switching.[12–14] Tuning this compensation temperature is a challenging prospect, as it is an outcome of the competing interactions between Sm- and Fe- sublattices. Changing the particle size by going down to the nanoscale and thereby changing the relative population of surface spins could possibly be exploited for tuning the compensation temperature.

Reducing the particle size in multiferroic/ferroelectric (FE) oxides can also affect the ferroelectric properties of the material. To ascertain the grain-size effects in such systems, the modified Ginzburg-Landau-Devonshire (GLD) phenomenological model has been formulated for classical[15,16] as well as improper[17] ferroelectrics. To aid the reader, a brief discussion of the GLD model is presented within the context of the current investigation.

We begin by briefly discussing the size reduction effect within the GLD model in the classical/proper ferroelectric $BaTiO_3$. A progressive decrease in tetragonal distortion, heat of transition, Curie temperature, and relative dielectric constant has been observed in dense $BaTiO_3$ ceramics with grain size decreasing from 1200 to 50 nm. The observations in dielectric depression has been attributed to the combination of the intrinsic size effect and of the size induced grain boundary changes.[18] As a characteristic feature at the nanoscale (< 100nm) intrinsic surface stress (ISS) in mono-domain grains is homogeneous. With reducing particle size, its contribution to the surface energy becomes comparable to and can even exceed the bulk energy.[19] Properties of dense nano-ceramic classical ferroelectrics are successfully described by their modified GLD model[15] and are in good agreement with the experimental data.[18,20,21] Here, the effects of increasing ISS with the particle size reduction assimilate those due to temperature increase viz., reduction of spontaneous polarization and the stability of the FE state.[22] In $BaTiO_3$ ceramics while for the larger grains, the stress developed upon the FE transition gets minimized by twinning in smaller particles the compressive ISS remains unrelieved, and tends to suppress the tetragonal deformation back toward the cubic state. Extrapolation of this suppression indicates a critical particle size (~20 nm) for the disappearance of FE in BTO nanoceramics.[18,20,21]

The modified GLD framework for improper FE's (incipient[19,23] and secondary ferroics,[17] with both $M$ and $P$ orders, mainly the type-II multiferroics such as SFO) incorporates the ISS, magneto- and electro-striction, as well as couples piezoelectric and piezomagnetic effects in the free energy of (essentially monodomain) nanoparticles.[17] Here, particle size reduction below ~50 nm substantially increases the surface energy vis-à-vis bulk's contribution. Enhanced ISS allied with growing (~1/grain-size) built-in fields and magneto-electric coupling, along with the restricted/confined dimensionality of the homogeneous order parameters in mono/aligned domain (nano-rods/spheres) is predicted to manifest a more prominent FE state. In particular, under favourable conditions,[17] ISS here is shown to increase $T_C$ and even induce the ordered state(s) below optimal nanosize in incipient ferroics, where the same may be elusive at larger particle sizes. Moreover, due to the striction- and ME-emergent secondary FE state, here a "critical" particle size signalling the disappearance of FE is not expected. Rather, the striction/ME effects grow with the particle size reduction. The documented enhancement of FE in Rochelle salt nanorods,[24] dramatically higher ME coefficients in epitaxially oriented $BiFeO_3$ films on a $SrTiO_3$ substrate[25] and observation of room temperature magnetism in nanospheres of $CeO_2$, $Al_2O_3$, ZnO etc.[26] all testify to the contrasting effects of grain size reduction in incipient vs. classical/proper ferroics.

In the present work, we have investigated the structural, magnetic, thermal, and dielectric properties of micro and nano-sized particles of SFO, synthesized using wet chemical route. In contrast to the literature-studies on SFO, we observed (i) significantly enhanced compensation temperature (transition temperature) (ii) irreversible remnance at zero magnetic field at low temperature and (iii) coupling of the antiferromagnetic-ferroelectric (AFM-FE) transitions. We also observe grains and defects within the nanoparticles via HRTEM, contesting the long-believed single crystalline nature of individual nanoparticles. We attribute the observed unique properties of the SFO nanoparticles to their core and shell structure and the associated surface/interfacial anisotropies. The dielectric data is also discussed in the context of previously reported theoretical work.

## II. EXPERIMENT

### 1. Sample preparation

SmFeO$_3$ nanoparticles were synthesized using a similar wet chemical route combined with post synthesis annealing, as reported earlier.[27] It involved a reaction of stoichiometric amounts of Sm(NO$_3$)$_3$.6H$_2$O and Fe(NO$_3$)$_3$.9H$_2$O in nitric acid with D L-tartaric acid, used as a complexing agent. The sol was heated at 353 K to form a gel-like precipitate. The gel was then heated in an oven at 423 K for 4 h. Samples were annealed at 973 K (SFO-1) and 1523 K (SFO-2) for 4 h each. The annealed powders were washed several times in Milli-Q water and ethanol before complete drying.

### 2. Characterization

Transmission electron microscopy (TEM) and high resolution TEM (HRTEM) analysis of synthesized powder samples (drop-casted on a copper grid, after dispersing in ethanol) of SFO-1 and SFO-2 particles were carried out using JEOL JEM – ARM200F microscope equipped with Schottky field emission gun. The energy dispersive spectroscopy (EDS) analysis was performed using EDS module attached to the instrument. The room temperature and high temperature X-ray diffraction patterns of the powder samples were collected in air using a Bruker AXS D8 ADVANCE diffractometer. The lattice parameters were obtained by Rietveld refinement using the software FullProf Suite (version July 2016). Magnetic measurements were carried out using the Vibrating Sample Magnetometer (VSM) attachment of a Physical Property Measurement System (PPMS) from Quantum Design. For low-temperature measurements, powders were filled in a polypropylene capsule that was mounted on the brass half-tube sample holder of the PPMS. The samples were centered at room temperature under a magnetic field of 10 mT. During the measurements, auto-centering of the sample was performed at 20 K steps to account for the thermal expansion/contraction of the sample holder. The measurements were performed in both zero-field-cool (ZFC) and field-cool (FC) protocols from $T$ = 310 K down to $T$ = 2 K. High temperature magnetic measurements were performed using the oven attachment of the VSM-PPMS in 300 to 750 K temperature range under high vacuum. Pellets were made out of powder samples weighing roughly 20-30 mg by applying pressure of 10 tons (diameter of the die used was 10 mm and the pellet was sintered at 673 K for 4 hours). The pelletized samples were mechanically anchored to the sample holder. The sample was loaded in the VSM oven under zero-field in ambient condition and the measurements were carried out while heating under an applied magnetic field of 0.1 T. Dielectric measurements were carried out on pellet samples (after gold-plating their flat surfaces) using an Alpha-A high-performance frequency analyzer (NOVO-Control Technologies) and a homemade oven, with isothermal control of $\Delta T \leq 0.5$ °C for each frequency-scan. Specific heat thermographs were collected on the $STAR^e$ DSC-1 (Differential Scanning Calorimeter, Mettler-Toledo) at the warm up ramp rate of 1 °C/min.

## III. RESULTS AND DISCUSSION

## IV. HRTEM

Figure 1 (a) shows low resolution TEM images of as prepared SFO-1 nanoparticles, which show that SFO-1 exhibits nearly spherical morphology with an average particle size ≈ 55 ± 5 nm. Interestingly, HRTEM analysis of SFO-1 nanoparticles shows presence of twinned domain structure (Fig. 1($a_1$)). This twinning is closely associated with ferroelectric-magnetic behavior of the nanoparticles, as reported in other perovskites.[28,29] The nanoparticles exhibit screw dislocations and stacking faults (Fig. 1($a_2$) and ($a_3$)). We note that dislocations and stacking faults are observed in proximity of the nanoparticles' surface, as seen in Fig. 1$a_3$. These defects contribute to the overall behavior of the nanoparticles. Figures 1($a_4$) and 1($a_5$) show the arrangement of Sm atomic-columns and Fe atomic-columns within the lattice. Clear grain boundaries can be observed in Fig. 1($a_3$) and 1($a_4$), showing more surface defects in the case of nano SFO-1, which could contribute towards extra local distortion.[30] It is noteworthy that this ordering is only stabilized at nanoscale in SFO-1. This gives rise to a peculiar magnetic behavior of this compound, which can be interpreted on the basis of antiferromagnetic core and ferromagnetic shell model. Here, the dynamics of antiferromagnetic spins within the nanoparticles and ferromagnetic unpinned spins at the boundary (FM-shell) of the nanoparticles play significant roles. This observation is important, since for long time it was believed that nanoparticles were perfect single crystals, except at the surfaces, but the occurrence of grains can help understand the properties of nanoparticles in new ways.

The low resolution TEM micrograph of SFO-2 shows average particle size ≈ 500 ± 5 nm (Fig. 1($c_1$)). These particles also exhibit twinned domains.[28,29] However, in SFO-1 these structures are distorted compared to SFO-2, indicating a major contribution from nano-size effect, maintaining the homogeneity in the domains. From Fig. 1($c_2$) and ($c_3$) it could be seen that the defects in SFO-2 are fewer compared to SFO-1, which can be attributed to high-temperature preparation process, which imposes better crystallinity. The improved crystalline nature of SFO-2 has also been revealed from the atomic-column arrangement for Sm and Fe atoms (see Figs. 1($c_4$) and 1($c_5$)). The EDS analysis in both the cases (SFO-1 and SFO-2) shows that elemental composition of Sm, Fe, and O within the nanoparticles (Figs. 1($b_1$-$b_4$) and bulk particles (Figs. 1($d_1$-$d_4$)) is nearly uniform. The quantification of peaks for SFO-1 and SFO-2 give the ratio of Sm:Fe:O as 1:1:3, which matches with the values obtained from EDS analysis of the nanoparticles, as shown in Figs. 1($b_4$) and 1($d_4$). Figure 1 (e) and (f) show the particle size distribution of SFO-1 and SFO-2 particles.

2. XRD

Figure 2 illustrates different bond lengths of Fe-O and Sm-O in $FeO_6$ octahedron and $SmO_{12}$ dodecahedron, in the orthogonal Pnma setting of the system, derived from the Rietveld refinement (The Rietveld-refined plots of X-ray diffraction data and derived structural parameters are shown in Supplemental Material Fig. (S1) and Table (S1)).[31] It is important to note here that the particle-sizes calculated from XRD using the Scherer formula turned out to be much smaller than that found using the TEM. For the reason that when the sample is annealed at higher temperature, not necessarily all the particles are grown into big particles; there is possibility of agglomeration of various crystallites into one grain /particle. Hence, the single particle (as seen by TEM) obtained after annealing at higher temperature may contain various crystallites, the SFO-1 and SFO-2 particles are not necessarily single structural domains.

The lattice parameters for the ~500 nm (SFO-2) particles are comparable to those reported by Maslen et al.,[32] while values for the ~55 nm (SFO-1) particles are noticeably different than those of the single crystal. The distortion emergent due to the particle-size reduction is further explained using the electron density plots of the samples, obtained from Rietveld refinement of the data. Two different planes of interest with appropriate intercept have been plotted, and shown in Fig. 3 for the SFO-1 and SFO-2 samples. The electron density (ED) maps play significant role in understanding the interactions at the atomic level. Inverse Fourier transformation of the structure factors $F_{hkl}$ obtained from the Rietveld refinement gives the electron density $\rho(x, y, z)$ as:[33]

$$\rho(x,y,z) = \sum_{hkl} \frac{F_{hkl} \cdot e^{\{-2\pi i(hx+ky+lz)\}}}{V} \quad (2)$$

Where (h k l) are the Miller indices and V is the volume of the unit cell.

First row in Fig. 3 shows the ED on the *xz*-plane taken at *y*-intercept of 0, and row 2 illustrates the sections of the ED in the *yz*-plane, taken at an *x*-intercept of 0.5, for SFO-1 and SFO-2. In the leftmost column in Fig. 3, the corresponding section of the unit cell is shown. In row 1, the sections are taken such that the electron density in the equatorial plane of the $FeO_6$ octahedra is evident. In this plane, the Fe-O2 bonds are visible and appear to form a rhombus comprised of Fe-atom at the center and oxygen atoms at the vertices (red solid rhombus drawn as a guide to eye). For SFO-2 the oxygen atoms connected along one diagonal have moved away from the Fe atom. The oxygen atoms connected along the opposite diagonal have moved closer to the Fe-atom at the center of the octahedra. As the particle-size is reduced to ~ 55 nm (SFO-1), the equatorial plane of the $FeO_6$ octahedra appears to change its shape from a rhombus of unequal diagonal lengths to one of almost-equal diagonal-lengths. Deduced Fe-O bond lengths shown in Table 1 also confirm this observation, as well as the observed reduced ED around Fe-atoms in the case of SFO-2. This indicates that Fe-atom has moved down in the '+*b*' direction in the case of SFO-1.

From row 2 we observe that in SFO-1 (as compared to SFO-2), the ED near Fe atoms decreases. The angle Fe-O1-Fe is 146.5° for SFO-1 and increases with the increase in particle-size to 148.4° for SFO-2. It is also noted here that the shape of electron density around Sm-ions is more anisotropic in the case of SFO-1. This investigation of two different particle sizes of SFO further adds in understanding the observed changes in magnetic and dielectric properties of the SFO nanoparticles of 55 nm.

3.     **Magnetic Properties**

Before presenting our data, let us briefly summarize various magnetic transitions previously observed in the temperature dependent magnetization of SFO single-crystal and ceramic samples. The Fe-spins order antiferromagnetically (AFM) below $T_N$ = 670 K,[8,34] with a weakly-canted ferromagnetic (FM) moment along the *c*-axis, which causes the *c*-axis susceptibility to exhibit a FM-like increase below $T_N$. In contrast, the *a*-axis susceptibility shows a small anomaly at $T_N$. Upon cooling the sample below $T_N$, near 470 K, the weak ferromagnetic (WFM) moment associated with the Fe-spins starts reorienting towards *a*-axis, which causes the *c*-axis susceptibility to decrease rapidly below this temperature, and simultaneously, the *a*-axis susceptibility to rise sharply. This temperature is, therefore, referred to as the spin-reorientation temperature ($T_{SR}$). At cryogenic temperatures, the magnetization of SFO exhibits a spontaneous reversal. This reversal is attributed to the compensation of WFM of the Fe-sublattice by antiparallel alignment of the $Sm^{3+}$ moments. Precise temperature ($T_S$) below which the $Sm^{3+}$ spins undergo long-range ordering is not known. However, from the bulk susceptibility there are indications that below about $T \sim 140$ K, $Sm^{3+}$ moments begin to order antiparallel to the Fe-moments. This antiparallel arrangement is presumably favored by the antiferromagnetic *f-d* exchange between the two sublattices.

## 3.1    Susceptibility Measurements

Magnetic susceptibility of samples SFO-1 and SFO-2 in the temperature range from 300 to 750 K is shown in Figs. 4(a) and 4(b). Rapid increase in the susceptibility around $T = 670$ K for both the samples coincides with the AFM ordering temperature of the Fe-sublattice. This increase is consistent with the fact that the spins in the AFM phase are canted, giving rise to WFM. The second anomaly near 470 K coincides with the spin-reorientation temperature ($T_{SR}$), below which the WFM moment of the Fe-sublattice rotates as discussed above. Temperatures $T_N$ and $T_{SR}$ for the two samples are in agreement with the previous reports on single crystal samples.[11,34] Within our measurement accuracy of about ±10 K, which is due to the high vacuum in the sample chamber and weak thermal-link between the sample and the thermometer, no appreciable change in either $T_N$ or $T_{SR}$ could be recorded between the two samples. There are, however, subtle differences in the temperature variations of their susceptibilities shown in Fig. 4. Namely, the magnetic transitions (both at $T_N$ and $T_{SR}$) in SFO-1 are not as sharp as in SFO-2; and the behaviour in the temperature range from $T_{SR}$ to $T_N$ differs slightly between the two samples. While in SFO-2, susceptibility decreases monotonically with increasing temperature up to $T_N$; in SFO-1 there is an intermittent rise of the FW curve above 550 K, resulting in a shallow peak just below $T_N$.

Additionally, it is observed that the magnetization of SFO-1 is suppressed with respect to SFO-2, measured under the same field and temperature conditions. As shown in the next section, magnetization isotherms in this temperature range also show consistently smaller magnetization values for the SFO-1 sample, up to the highest applied fields.

Since the temperature variation of susceptibility in SFO-2 is in accordance with the previous single crystal reports, the different behaviour of SFO-1 can be attributed to its smaller particle size. Magnetic nanoparticles of a canted antiferromagnet often exhibit a core and shell nanostructure, due to the uncompensated spins and crystalline imperfections near the surface of the particle. SFO is a canted antiferromagnet and the TEM images of SFO-1 suggest the presence of core and shell type of structure. Magnetic properties of core and shell nanostructures often show considerable departure from their bulk counterparts, due to the existance of anisotropies, resulted from the under-coordinated surface and interfacial-defects.[35–37] Since the overall magnetization of SFO-1 is considerably suppressed relative to that of SFO-2, it indicates that the surface spins have their net magnetization oriented antiparallel to the weak-ferromagnetic (WFM) moment of the core, reducing the overall magnetization. Magnetization-reduction due to the pinning of the surface spins with net magnetic moment pointing antiparallel to the core, has been reported previously for several ferrimagnetic nanoparticles.[35,37] The observed behaviour of SFO-1 is mainly attributed to the interplay of spins in the core and the shell regions, along with these, there are minor contributions from defects in interface region and also from the structural domains seen in the TEM images. It is hypothesized that upon changing the temperature across $T_{SR}$, the interfacial-spins remain randomly pinned, making the anomaly at $T_{SR}$ in the core and shell particles relatively broad. However, beyond a certain high temperature, the thermal energy available to these spins will overwhelm the pinning barrier, rendering them free to rotate parallel to the applied field, resulting in the observed intermittent-increase of the susceptibility above $T = 550$ K. The final decrease upon increasing the temperature of the sample close to $T_N$ is due to a complete loss of the spins long-range ordering, as the sample turns paramagnetic. The presence of pinned interfacial spins is also manifested in the isothermal magnetization presented later. We shall now compare the low-temperature magnetization behaviour of the two samples measured under an applied magnetic field of 1 mT. Low-field susceptibility of the two samples is shown in Figs. 4(c) and 4(d). The measurements were carried out in the temperature range from 300 K down to 2 K under the zero-field-cooling (ZFC) condition. In both samples, upon cooling below 300 K the susceptibility increases slowly down to about 140 K and decreases thereafter. This temperature (marked as $T_S$) coincides with the temperature where the $Sm^{3+}$ moments start to align antiparallel to the Fe spins, as reported previously for the single crystal.[8] The manner of decrease of the susceptibility below $T_S$ is, however, different for the two samples. In SFO-1, it starts showing considerable decrease right below $T_S$; on the other hand in SFO-2, it decreases slowly down to 50 K, followed by a sharp drop upon cooling below 50 K.

In the low temperature region, both samples exhibit magnetization-reversal below a characteristic compensation temperature (marked as $T^*$), where the ordered moment on the $Sm^{3+}$ sublattice exactly cancels out the ordered moment on the Fe sublattice. $T^*$ is found to be 4 K for SFO-2 and 22 K for SFO-1 (see insets in Figs. 4(c) and 4(d), respectively). The low-field temperature-variation of the susceptibility of SFO-2 shows a good agreement with the single crystal data.[34] The compensation temperature of SFO-2 ($T^* \sim 4$ K) is also in good agreement with that of the single crystal. In SFO-1, on the other hand, not only the $\chi(T)$ behaviour is different, compensation temperature is also considerably enhanced. These changes are consequent to the reduced particle size and the core and shell nanostructure of SFO-1. During the compensation process, as the $Sm^{3+}$ moment grows to cancel the WFM of the Fe spins in the core, the surface spins aid the $Sm^{3+}$ moments, since they are pinned with their net moment aligned antiparallel to the core's WFM moment, as discussed above. In this scenario, the magnetization compensation in the nanoparticle sample is expected to take place at temperatures higher than that in the bulk sample, as is found to be the case here.

### 3.2     Isothermal Magnetic Measurements

To study the effect of particle size further, we carried out *M-H* measurements at various temperatures on both our samples. SFO-2 is studied as the control sample, which has the properties analogous to that of the bulk. Figures 5 (a-d) show the magnetic hysteresis loop of samples SFO-1 and SFO-2 for *T* = 2K, 70 K, 310 K, and 650 K. Insets show the zoomed-in view of the magnetization plots near the origin. We also collected data at intermediate temperatures of 400 K and 550 K, which are not shown here as they do not exhibit any qualitative difference from the data at 310 K. The *M-H* loop parameters, namely, magnetization under the highest applied field ($M_{H0}$), remnant magnetization ($M_r$), coercivity ($H_c$), and shift on field-axis $H_{shift}(T)$ obtained from these measurements are shown in Table 1.

Table1. Derived magnetic parameters.

| T | $H_{shift}(T)$ | | $H_C(T)$ | | $M_{H0}(T)$ | | $M_r(T)$ | |
|---|---|---|---|---|---|---|---|---|
| | SFO-1 | SFO-2 | SFO-1 | SFO-2 | SFO-1 | SFO-2 | SFO-1 | SFO-2 |
| **2K** | 0.615 | 0.475 | 0.105 | 0.025 | 736 | 745 | -56 | -46 |
| **70K** | -0.455 | -2.200 | 0.115 | 0.600 | 371 | 367 | 38 | 106 |
| **310K** | 0.000 | 0.000 | 0.900 | 0.680 | 349 | 420 | 58 | 110 |
| **400K\*** | 0.003 | 0.001 | 0.392 | 0.273 | 160 | 223 | 64 | 106 |
| **500K\*** | 0.002 | 0.002 | 0.360 | 0.228 | 152 | 202 | 55 | 97 |
| **650K** | 0.000 | 0.000 | 0.072 | 0.258 | 86 | 125 | 9 | 41 |

*\*Plots not shown in the manuscript.*
*\*Highest applied field ($M_{H0}$) is different for different temperatures.*

As shown in Fig. 5, for both the samples SFO-1 and SFO-2, the *M-H* plots at low-temperatures (panels (a) and (b)) show characteristically different behaviour from those at high-temperatures (panels (c) and (d)). The low-temperature plots (*T* = 2 K and 70 K) are characterized by their shifted *M-H* loops. In contrast, the *M-H* loops at high temperatures are symmetric and well-formed, as in typical ferromagnets, but without any signs of saturation up to the highest applied field. The data at low temperatures is recorded below $T_s = 140$ K, where the Sm-sublattice magnetization starts growing upon cooling. The *M-H* loops at these temperatures are thus strongly reflective of the antiferromagnetic coupling between the Sm/Fe sublattices, and the strong single-ion anisotropy of the Sm moments. The electron density maps at 300 K suggest that the 4*f* electron cloud surrounding the $Sm^{3+}$ ion in SFO is anisotropic in shape. Deviations of the charge density of 4f electrons of rare-earths from a spherical symmetry arises due to the crystal field splitting of the spin-orbit coupled lowest *J*-multiplet (*J* being the total angular momentum for a non-zero orbital angular momentum). This is well-known to impart a strong single-ion anisotropy to the rare-earth moment. At low-temperatures (below $T_S$) therefore, unless the applied field exceeds either the anisotropy field or the field equivalent to the *f-d* exchange, the magnetization is expected to remain linear, as is found to be the case here.

The shift of the *M-H* loops at 2 K and 70 K is huge. Typically, *M-H* loop shift in the core and shell nanostructures is attributed to the exchange-bias effect. However, here we found equally pronounced shifts even for our SFO-2 sample, which suggests that the core and shell morphology alone is not enough to explain the shifts. Moreover, since the hysteresis loops at low temperatures are not saturated i.e., those are minor loops, the observed shift can not be unambigously attributed to the exchange bias effect.[38] Large shift of the *M-H* loop for the two samples at low temperatures could be arising from the complex interplay of magneto-crystalline anisotropy and the *f-d* exchange,which needs to be investigated further in detail.

We discuss next the *M-H* plots at higher temperatures where the role of $Sm^{3+}$ moments and the resulting crystal-field-derived single-ion anisotropy are not dominant. Accordingly, the *M-H* plots at higher temperatures are far more symmetrical and well-formed at 310 K and 650 K, as shown in Figs. 5(c) and 5(d). Both SFO-1 and SFO-2 show the expected polycrystalline average behavior. In SFO-1 however, two new features, not present in the magnetization of SFO-2, are also noted: first, the jump or discontinuous change in *M-H* near $H = 0$ and second, the smaller magnetization at all temperature/field values compared to SFO-2. Since the surface spins, due to the strong surface anisotropy, are pinned with a net magnetization aligned antiparallel to the core, the overall magnetization in nanoparticles (SFO-1) is suppressed, compared to the bulk sample (SFO-2). At the 310 K, the coercivity of SFO-1 is higher than that of SFO-2, which can be attributed to their core and shell nanostructure. In a previous single crystal study, the coercive field at 300 K for $H \parallel a$-axis is reported to be less than 50 Oe.[34]

The jump in *M*(*H*) near $H = 0$ in SFO-1(Figs. 5(c) and 5(d)) is clearly discernible near room temperature but gradually weakens upon heating, disappearing completely at the Néel temperature. A similar behaviour has been previously observed in $YFeO_3$ nanoparticles.[39] Previous reports pointed out the presence of surface/interface anisotropies in the core and shell nanostructures.[35,37,40] Typically, the unidirectional anisotropy at the core and shell interface is expected to be weaker than at (or near) the surface, because the spins at (or near) the surface are significantly under-coordinated, whereas spins in the interface region experience crystalline defects. As the strength of the applied field exceeds the anisotropy fields pinning the spins, a discontinuous change in the magnetization results due to depinning. The jump in *M-H* of SFO-1 near $H = 0$ is probably due to the depinning of the interfacial spins. Additionaly, a weak pinning of the magnetic domian walls due to the structural domains will also afect to the *M*(*H*) loop near $H = 0$.

We also examined the behavior of the coercive field for the two samples as a function of temperature. If one looks at the temperature dependence of the coercive field of SFO-1 (see, Supplemental Material Table S2),[31] it decreases monotonically with increasing temperatures as expected, because increasing the thermal energy facilitates the domain-wall motion. However, that of SFO-2 appears to be anomalous, because in this case the coercive field increases upon increasing the temperature from 500 to 650 K. This behaviour has previously been found in some Sm-Co based high-temperature magnets. Though, far from fully understood, it is generally argued that this behaviour is related to the multi-phasic nature of these materials; having magnetic phases in close proximity with different domain-wall energy densities.[41] Also since the rare-earth ions in orthoferrites are paramagnetic and can be magnetized by the Fe moments. The nature of slope is similar for SFO-1 and SFO-2 samples in Figs. 4 (a) and (b), indicating similar Sm densities. While, the contribution of Fe moment (remnance) is greatly reduced in SFO-1 sample (Fig. 5). This is consistent with the change of bonding in SFO-1 from the XRD analysis. Hence the Fe-Sm exchange interaction, which is responsible for magnetization of Sm, is also expected to be different, which will also contribute to the observed increase in the compensation temperature. It will be interesting in future to investigate the origin of the observed anomalous behavior in micrometer particle size samples of $SmFeO_3$.

4.  **Dielectric Measurements**

From the magneto-electric (ME) perspective, we have seen (SFO-1) and extracted (SFO-2) signature anomalies at $T_{SR}$ and $T_N$ from their dielectric measurements. Complementing the slope-break feature seen in supplementary Fig. S2(a),[31] the spin reorientation temperature also marks the minima of conduction and losses in SFO-1 (Fig. 6(a)), with their similar retracing behaviour from below to $T_{SR}$ and above. Parametric evolution (away from $T_{SR}$) of the charge barrier-activation and dipolar relaxation thus creates almost similar co-regression between the conduction and loss characters, of the *a*- and *c*-axis spin-oriented phases. Hidden signature of $T_{SR}$ in SFO-2 (~500 nm) gets revealed by fitting the full Jonscher[42,43] function {$\sigma_T(f) = \sigma_{dc} + A f^s = \sigma_{dc}[1+(\omega/\omega_h)^s]$} onto its conductivity spectra, shown in Fig. 6(b). Exponent *s* here represents the degree of interaction between the charge carriers and the lattice, and is related to the dimensionality of the conduction pathways;[44] whereas $\omega_h$ is the charge-hopping frequency. Despite relatively small changes of index *s*, its obtained qualitative behaviour is well clear of the uncertainties (Fig. 6(c), bottom plot). The observed decreasing $s(T)$ due to the correlated barrier hopping (CBH)[45,46] below $T_{SR}$ reverses to the increasing trend above, attributed to the small polaron tunneling (SPT).[47,48] Sharp behaviour-breaks in $A(T)$ and $s(T)$ speed-up the dispersion ($\Delta(d\ln\omega_c/dT)|_{T_{SR}} \sim 14\%$), of the characteristic dc/ac crossover frequency, given by $\omega_c = \omega_h$ (c.f., the locus of $\sigma_c = 2\sigma_{dc}$ traced vs. temperature, Fig. 6(b)). Moreover, two different Arrhenic behaviors of the fitted $\sigma_{dc}$ vs. $1/T$ (not shown) evidence a step-up in the activation energy, $\Delta E_a|_{T_{SR}} \sim 12\%$. Switching of 3-D[44] conduction-mechanism concurs the retreat of the 'FE' state above $T_{SR}$, into a relaxor-like short-range order below, latter recently reported[7] in SFO-1. Electrically, spin-reorientation thus registers in SFO-2 rather subtly (albeit profoundly, in magneto-sensitive conductivity character), versus its anomalous signature-absence in the single crystal.[11,34]

The indirect $T_{SR}$-anomaly revealed in SFO-2 (~500 nm grains) via the parametric $\omega$-dependence over the RF-range ($\leq$ 100 kHz, Fig. 6(b)) should also manifest in SFO-1 (~55 nm grains), but over the microwave range (> 1 MHz, not probed here); a hint of the corresponding power-law regime is marginally observable at the high-frequency-end of the conductivity-isotherms, recently reported in a 50-60 nm SFO sample.[49] This circumstance is due to the inverse relation between the characteristic frequency- and length-scales.[50,51] In turn, this also explains the detection (absence) of ($d\varepsilon'/dT$)-discontinuity at $T_{SR}$ in SFO-1 (SFO-2) over the RF-range (supplementary Fig. S2);[31] in SFO-2, the relatively large-valued activation/relaxation parameters $E_a$ and $\tau$, which determine the mostly polaronic $\varepsilon'(T)$, seem insensitive to their small steps $\Delta E_a(T_{SR})$ and $\Delta\tau(T_{SR})$.

Concerning the Jonscher analysis presented here for SFO-2 conductivity, we remark that (1) the highly-subdued extrinsic contributions (low-magnitudes of $\varepsilon$ and $\sigma$) are downshifted to lower frequencies; little influencing our fitting, (2) the intrinsic (intra-grain) contribution to $\sigma_{ac}$ manifest at the RF-range is fully accessed, and (3) dc-conduction regime is accessed by the data, and so $\sigma_{dc}$ is incorporated in the fitting procedure. Thus, the complete fittability of our full $\sigma$-isotherms (single dc- and ac-regimes) lends high accuracy to the evaluated parameters, and to the consignment of intrinsic conduction mechanisms based on their temperature dependencies.

To place our findings in perspective, note that a recent work[49] on a 50-60 nm SFO reported vastly higher dielectric constants (x10$^3$) and conductivities (x10$^4$) vis-à-vis those measured on our similarly-sized SFO-sample.[7] The extraneous effects of electrode-surface layer and grain-boundaries, which are strongly dependent on the post-synthesis sample-processing/preparation, radically increase these basic dielectric attributes.[52–54] This is particularly true e.g., in the case of qualitatively large difference between the inter- and intra-grain conductivities, of the so-called colossal dielectric constant (CDC) materials,[55] which can swamp the detection of weak electrical phenomena in their wake.[56] CDC character of the results by Sahoo et. al.[49] is amply evident by their multitude of data and fittings, even as the actual intra-grain $\sigma(\omega)$ is non-analysable (lying over $\geq O$ (MHz); besides the limited frequency-window accessed here, the data accuracy may be compromised due to the instrumental range-end). Moreover, non-realization of the dc-transport regime indicates the prevalence of extraneous effects over the usual (radio frequency, RF) range analysed.[49] The intrinsic nature of our data has been reported recently[9] and for conciseness and space-economy, here we consider only our most relevant data-representations and analyses.

Next, we observe small but clear non-dispersive peaks exactly at $T_N$ in the permittivity data of SFO-1 (Fig. 7(a)); a signature which is fairly reasonable to associate with the FE transition at $T_C$, as coupled to the concurrent AFM-$T_N$. Existence of the $\varepsilon'(T_N)$ peak-anomaly (hitherto unreported in the single crystal and bulk samples of SFO) actually ensures an appreciable ME-switchability of the polar state, as an essential feature of a multiferroic.[57–60] Nature of the $\varepsilon'$-peak anomaly here in SFO-1 (viz., markedly narrow $\Delta T$ and $\Delta \varepsilon$) is akin to the AFM-induced FE transition reported in doubly-doped Nickel Oxide (Li$_{0.05}$Ti$_{0.02}$Ni$_{0.93}$O)[61] and in single crystal Cupric Oxide (CuO),[62] high temperature multiferroics. The existence of polarization and piezoelectricity in single crystal SFO was first reported by Lee et. al.;[34] the lingering issue, however, has been the absence in the dielectric constant of any signature of FE, and the contested nature of its ME-coupling. Initially,[34] the FE was presented as brought about by the spin-canting, breaking the inversion symmetry and generating a weak electric dipole (via reverse D-M interaction). However, on grounds of group-theoretical analysis, this claim ($S_i \times S_j$-type ID-M polarization) was discounted by Johnson et. al. and asserted[63] rather in favor of the exchange-striction ($S_i \cdot S_j$-type polarization) as being the relevant mechanism for improper FE in SFO. Though more abundant electrical data are yet mandated to settle all the issues concerning magneto-electricity in SFO, the present results signify the (nano)structural lengthscale (≤ 60 nm), as favorable for the relevant investigations. Additionally, we observe knee-anomalies in the dielectric constant of the SFO-2 sample (Fig. 7(b)) some 30 K above $T_N$, which we attribute to the signture of an incipient/non-robust ferroelectricity. Although unrelated to the coupled-FE transition, it probably manifests the ME effect of the frustration-rooted AFM-correlations above $T_N$. These constitute the functionally important signatures of magneto-electric effect in the system and its significant particle-size dependence. By all accounts, the ME cross-coupling is evidently more prominent in SFO-1, highlighting its crucial functional advantage over both SFO-2 and the bulk counterparts.[11]

Since $\varepsilon'$ is the response parameter to an applied $E_{ac}$-field (which does not couple to the spins directly), the nature of the above $\varepsilon'$-anomalies should relate only to the electrical substructure of the grains. Typically, the FE domains have mesoscopic size scale $\sim O(10^2)$ nm.[64,65] Therefore, the ~55 nm grains of our SFO-1 are reckoned to comprise of lone FE domains each, whereas the ~500 nm SFO-2 grains must host some $\sim O(10)$ FE domains. The incipient FE, which is debatable in the single crystal and undetected in polycrystalline SFO, may be just emergent in SFO-2, via a broad precursor-anomaly in the dielectric constant above $T_N$ (Fig.7(b)). However, as discussed below, the much stronger intrinsic surface stress (ISS), together with the comparable (rather dominant) shell contribution vs. that of the core, can tip the FE character from being incipient in SFO-2 to an induced one, in the smaller sized SFO-1. Relative orientation of the lattice and the domains within a grain being uncorrelated, though the grains of SFO-1 themselves are randomly-oriented, the FE-domains singly hosted within them are less randomly aligned. While the 90° domain walls may be present in SFO-2, the non-polar mesoscopic grain boundary interface isolates the lone FE-domains (per grain) in SFO-1. The more uniform, untwinned FE-domains in SFO-1 are more prone to alignment under the internal $E_{ME}$ field, born upon the long-range AFM ordering. Under these circumstances, it is highly plausible for SFO-1 to stabilize a finite/switchable polarization upon AFM-$T_N$ and for SFO-2 to feature only an altered polarizability above AFM-$T_N$, which manifest as corresponding sharp and subtle magneto-electric signatures in Figs. 7(a) and 7(b) respectively.

Regarding theoretical support by a suitable modified-GLD framework, the present results may be qualitatively compared to some of the predicted general features of model calculations, available for particulate-shaped nanoparticles of secondary ferroics/multiferroics. It must be noted that the calculated[17,18] peculiar size effects are expected to optimally manifest in ≤ 50 nm nanorods (2D confined & 1D extended nanostructure) and less favorably (i.e., below shorter lengths and lower temperatures) in ≤ 5 nm nanospheres (3D confined & "0"D extended nanostructure). For secondary ferroics, model calculations are not available for thin films (1D confined & 2D extended nanostructure, with 'favourable conditions' expected to be least restrictive), to which the shell of our SFO-1 (~55 nm) homomorphically corresponds. This circumstance is evident by associating the AFM order here with the core ('bulk'); the Nèel temperature remains essentially the same in SFO-1 (55 nm) and SFO-2 (500 nm), whereas due to the dominant effects of the shell (closed "thin film") in SFO-1 (55 nm), the coupled FE-$T_C$ anomaly is directly 'detectable' in its dielectric constant (pending independent/supplemental material results).[31] Therefore, the "incipient ferroelectricity" which is latent in polycrystalline SmFeO$_3$ and faintly hinted here in SFO-2 (weak "thin film" like contribution, Fig. 7(b)) metamorphoses as an 'induced FE' state in SFO-1 (dominant surface/shell contribution, Fig. 7(a)). Moreover, the localization of the primary magnetic (allied electrical) order exclusively with the core (shell) and the doubly-strong ISS (due to the two, outer and inner surfaces) of the shell endows our nanostructure the character of a composite/bilayer (especially at smaller size), for which the 'extrinsically-rooted' ME coupling is well-documented to be rather huge,[66,67], due to their coupled magneto-striction and piezoelectricity. While the task of modelling the size-dependence of multiferroicity in thin-films/composite-bilayers is certainly beyond the scope of the present paper, it is highly desired that our results attract the interest of relevant theoretical groups to undertake the same. Nonetheless, based on the axial character (i.e., along the length as the extended-dimension) of polarization in the nanorod GLD model,[17,18,21] we expect the polar order here (localized in the mono-domain shell) to be aligned tangential to the longitudes. Such texture suppresses both the gradient- and depolarization-field terms in the free energy (which blur/smear the $T_C$ in bulk), as well as enhances the polarisation-switching (under an applied/ME $E_{ac}$ field), facilitating the FE-$T_C$ detection directly as the $\varepsilon'(T_N)$-anomaly (Fig. 7(a)). Contrasting nature of the $T_N/T_C$ anomaly observed in specific heat thermograms (supplemental material Fig. S3)[31] is consistent with the estimated assertion of single/multiple FE domains. The discrepancy between the transition temperatures observed in these thermograms and those marked in the dielectric permittivity (Fig. 7(a) and 7(b)) is presently unclear, and requires further investigation.

## IV. CONCLUSIONS

Two different sizes of SmFeO$_3$ nanoparticles varying by an order of magnitude i.e. ~55 nm and ~500 nm are studied. The detailed structural characterization is carried out to understand the dynamics of Fe- and Sm-sublattices. The high- and low-temperature magnetization studies to investigate any possible deviation from the various reported transitions in the ~55 nm particle, and high-temperature dielectric studies across the spin reorientation and antiferromagnetic transitions are performed. Magnetic signature of spin reorientation is evident in both the samples at 480 K, similar to that of the single crystal. Further, four new/yet-unreported signatures for ~55 nm nanoparticles are observed; (i) most significantly, the compensation temperature is achieved at 22 K, much higher than 4 K, reported for the single crystal and bulk SmFeO$_3$. The compensation due to the magnetization reversal of the Sm- and Fe-sublattices is observed at a much lower applied field (1 mT), compared to the studies on single crystals, which obtained switching at 0.01 T. (ii) anomalous behavior over 550 K to 630 K, driven by depinning of the interface spins and (iii) for ~55 nm sized SFO-1 sample, the spin-reorientation transition is directly discernible in its temperature-dependent dielectric data. For ~500 nm sample however, frequency-dependence of conductivity divulges that sub-$T_{SR}$ correlated barrier hopping (CBH) classical transport switches to small polaron tunneling (SPT) quantum transport above $T_{SR}$. In ~55nm particles, the signature of ME-coupled AFM-FE transition is directly observed in its dielectric constant. The induced mono-domains in ~55nm particles and incipient poly-domains in ~500nm particles are respectively consistent with the sharp peaks in the dielectric constant at $T_N$/$T_C$ in SFO-1 and the broad anomaly above $T_N$ in SFO-2. This endows a clear functional superiority to the smaller-grained SFO material for magneto-electric applications.

## ACKNOWLEDGMENTS

This work was carried out under the grant no. SR/WOS/-A/PS50/2012(G) availed by S.C. from Department of Science and Technology, Ministry of Science and Technology India and grant no. SR/NM/TP-13/2016 availed by S.S. A.M.A. acknowledges help with the dielectric and thermal data-collection from Suresh Bhardwaj and Amit Kumar Naiya.

**Supplementary information**: Reitveld refined X-ray diffraction data, derived structural parameters, dielectric, and specific heat measurements.

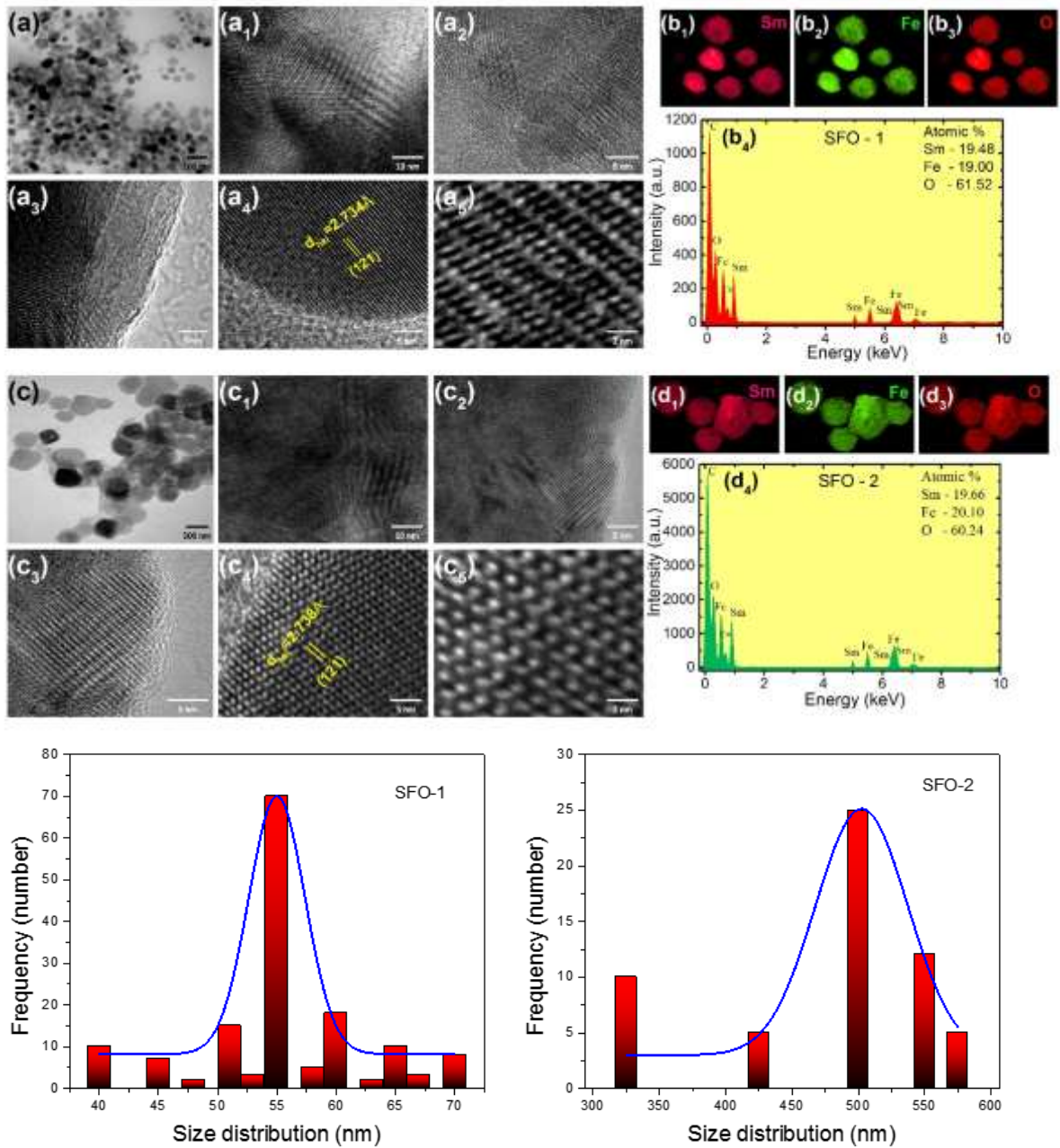

**FIG. 1**. TEM, HRTEM, and EDS analyses of SFO-1 and -2. (a) low resolution micrograph of SFO-1, ($a_1$) – ($a_3$) HRTEM micrographs of SFO-1 representing twinning domains, presence of defects, core – shell behavior respectively, ($a_4$) and ($a_5$) shows Sm and Fe atomic columns within $SmFeO_3$ lattice for SFO-1, ($b_1$) – ($b_4$) shows elemental mapping and EDS spectrum respectively for SFO-1. (c) low resolution micrograph of SFO -2, ($c_1$) – ($c_3$) HRTEM micrographs of SFO-2 representing distorted twinning domains and presence defects respectively, ($c_4$) and ($c_5$) shows atomic column arrangement for Sm and Fe atoms within the lattice, ($d_1$) – ($d_4$) represent elemental mapping and EDS spectrum of SFO-2 respectively and (e) and (f) show particle size distribution for SFO-1 and SFO-2 as obtained from TEM.

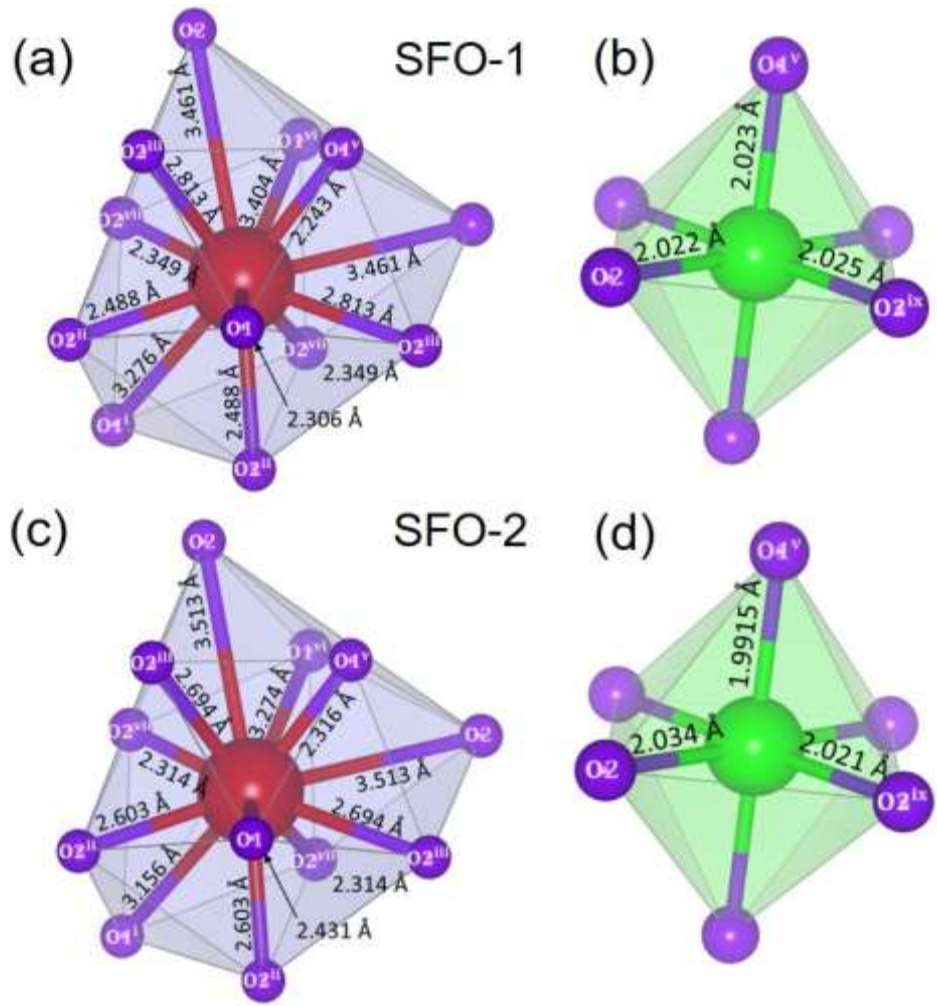

**FIG. 2**. Illustrating the Sm-O and Fe-O bond lengths. (a) $SmO_{12}$ dodecahedron of sample SFO-1(nano), (b) $FeO_6$ octahedron for sample SFO-1 (nano), (c) $SmO_{12}$ dodecahedron of sample SFO-2(micro) and (d) $FeO_6$ octahedron for sample SFO-2 (micro).

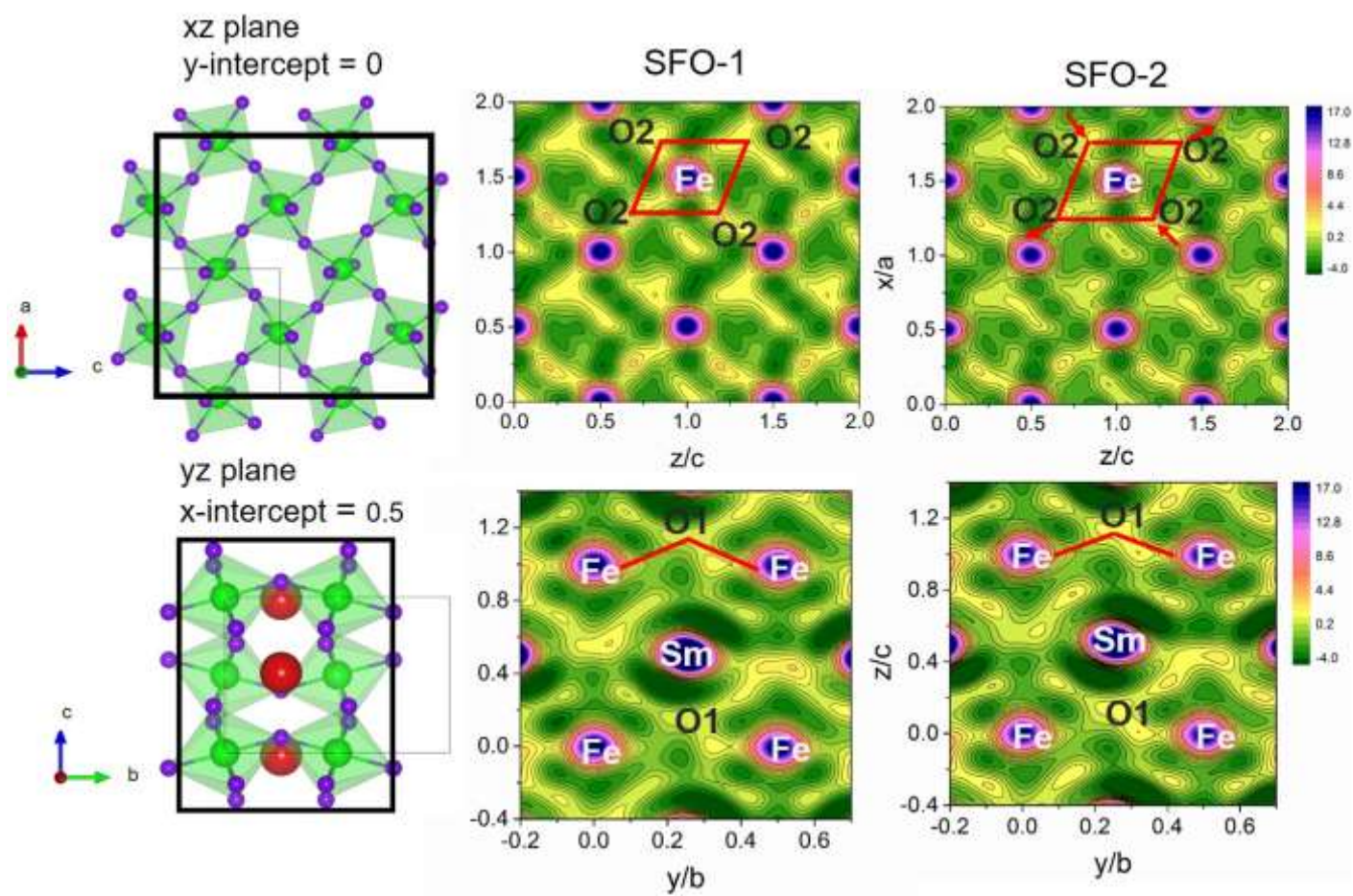

**FIG. 3.** Electron density maps obtained by Fourier transform of Rietveld refined data. The ED plots for samples SFO-1 and SFO-2 illustrating the change in Fe-O1-Fe angle and Fe-O2 bond lengths corresponding to nano and micro size particles.

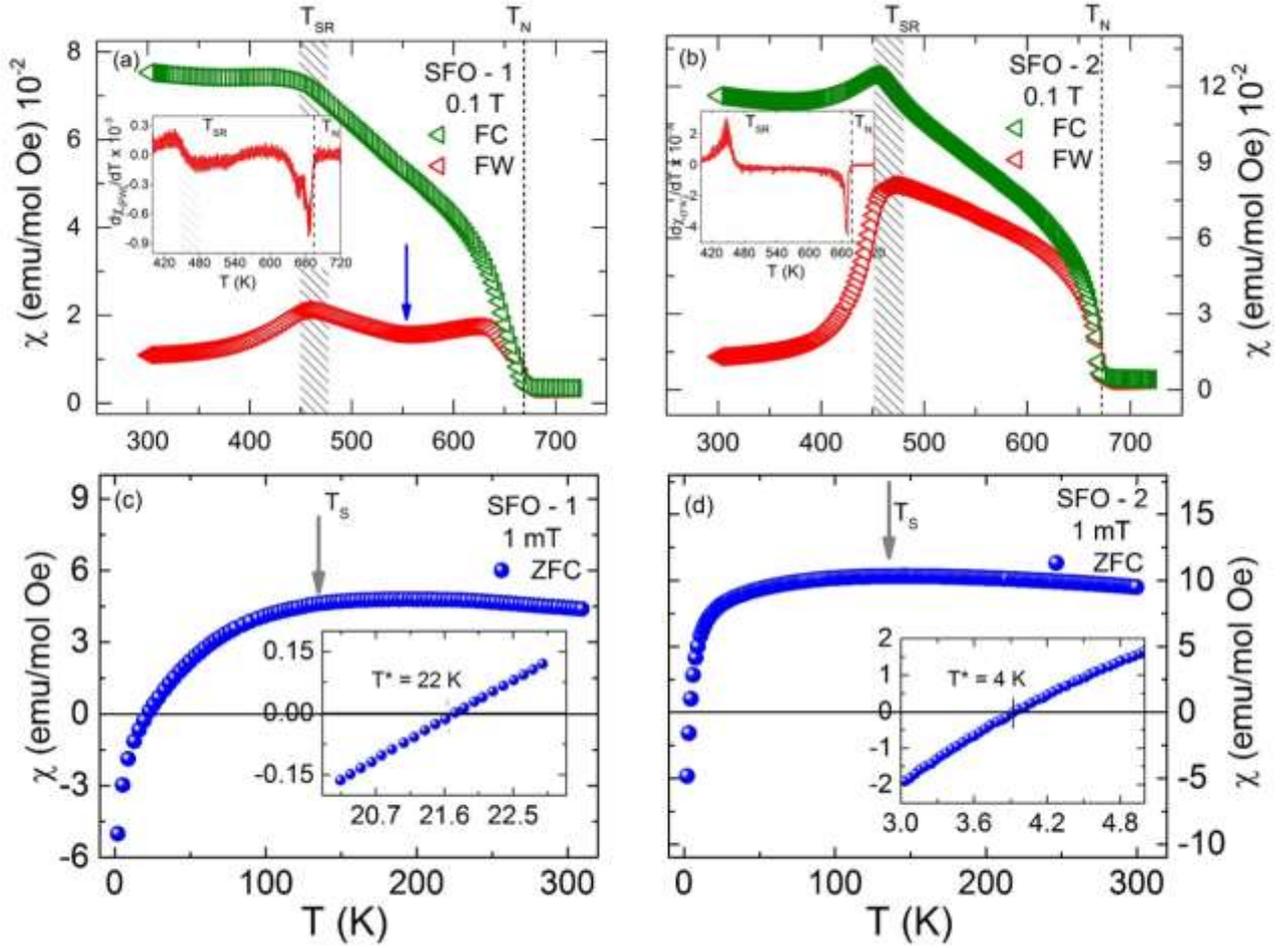

**FIG. 4.** High and low temperature dependent susceptibility curves showing various transition temperatures in SFO nano- and micro- particles: $\chi$-$T$ curve for temperature 300 K to 750 K for samples (a) SFO-1 and (b) SFO-2 respectively at magnetic field 0.1 T depicting $T_{SR}$, and $T_N$ (derivative of $\chi_{FW}$ in inset) for both the samples and an anomaly at ~630 K in case of SFO- 1. $\chi$-$T$ curve for temperature 2 K to 300 K for samples (c)SFO-1 and (d) SFO-2 respectively at magnetic field 1mT showing compensation temperature $T^* \sim$ 22 K for SFO-1 and ~4 K for SFO-2.

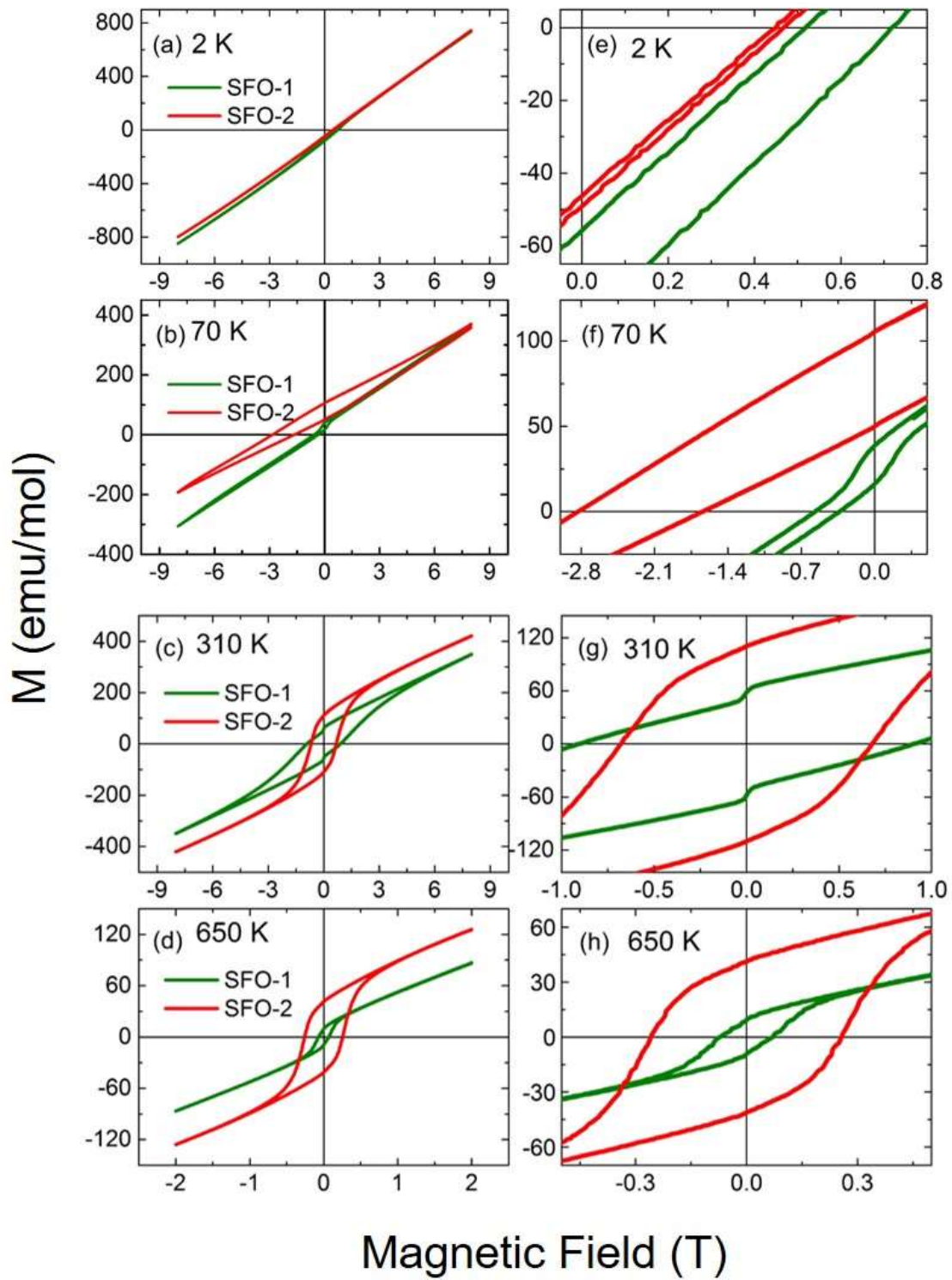

**FIG. 5**. Magnetic Hysteresis measurements for samples SFO-1 and SFO-2 at various temperatutres: (a) 2 K, (b)70 K, (c) 310 K and (d) 650 K. Figures (e) to (h) show zoomed-in view of the corresponding magnetization plots.

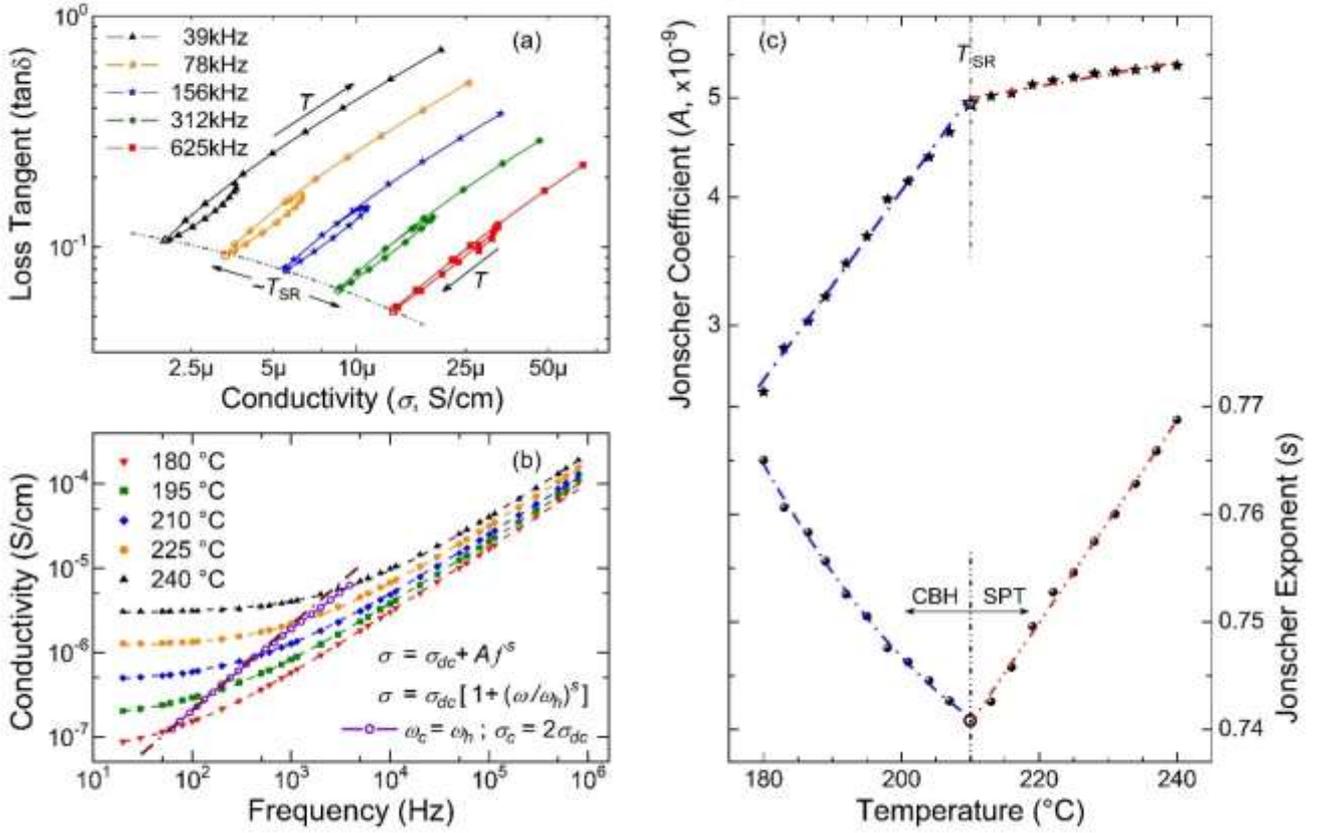

**FIG. 6**. (a) Co-regression of the activation (conductivity, $\sigma$) and relaxation (loss tangent, $\tan\delta$) behaviors and its near similarity above and below the spin-reorientation temperature in the SFO-1 (~55 nm) sample marks $T_{SR}$ as the turnback double-minima. (b) For SFO-2 (~500 nm), Jonscher-form-fitted conductivity isotherms at benchmark temperatures across the $T_{SR}$, along with the locus traced out by their dc/ac crossover ($\omega_c = \omega_h$, $\sigma_c = 2\sigma_{dc}$). (c) Jonscher-coefficient ($A(T)$, top) and power-law exponent ($s(T)$, bottom) determined from the empirical fits to the conductivity isotherms mark the spin-reorientation transition, via switching of the conduction behavior from classical correlated barrier hopping (CBH) below $T_{SR}$ to quantum small polaron tunneling (SPT) above. The behavior-breaks found at $T_{SR}$ in $A(T)$ and $s(T)$ (shown), and also in Arrhenic $\sigma_{dc}(T^{-1})$ (not shown) correspond to discontinuities in the dispersion of the dc/ac crossover $\Delta(d\ln\omega_c/dT)|_{T_{SR}} \sim 14\%$ (open circles in (b)) and in the activation energy $\Delta E_a|_{T_{SR}} \sim 12\%$.

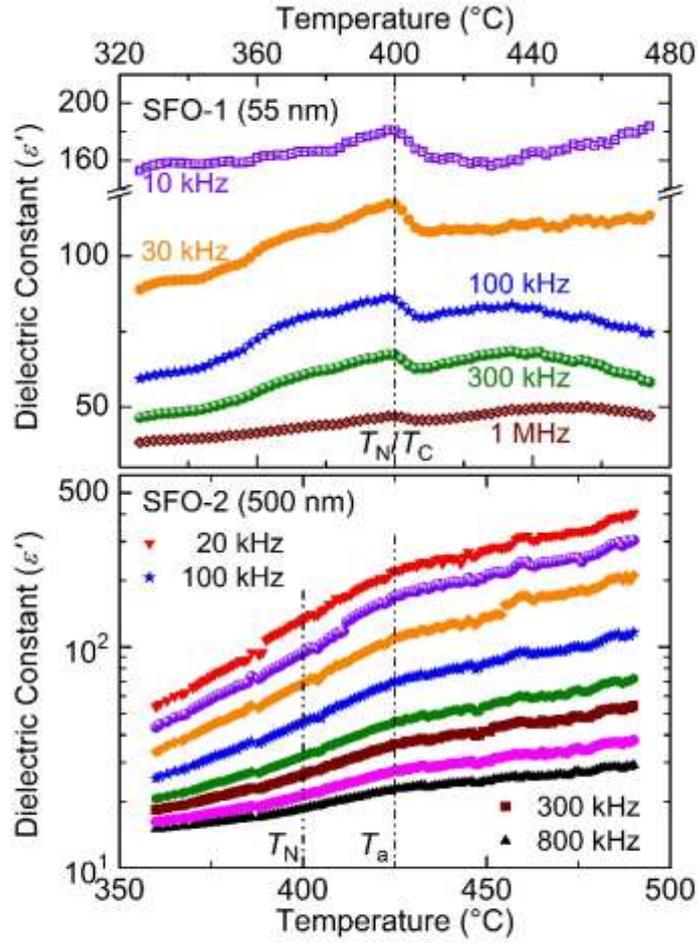

**FIG. 7.** Anomalies in the dielectric constant indicate respectively the emergence of magneto-electrically 'induced-FE' phase upon the AFM ordering ($T_C = T_N$, SFO-1, ~55 nm, top-panel) and the 'incipient-FE' signature above the AFM ordering ($T_a > T_N$, SFO-2, ~500 nm, bottom-panel). Contrasting nature of the anomalies is consistent with the single/multiple FE domains, reckoned as hosted by the nano-/micro-sized grains of the two specimens respectively.

# Supporting Information

# Nanosize effect: Enhanced compensation temperature and existence of magneto-dielectric coupling in SmFeO$_3$

*Smita Chaturvedi,[a*] Priyank Shyam,[a] Rabindranath Bag,[a] Mandar M. Shirolkar,[b] Jitender Kumar,[a] Harleen Kaur,[a] Surjeet Singh,[a,d] A.M. Awasthi,[c] and Sulabha Kulkarni[a*]*

[a]*Indian Institute of Science Education and Research, Pune, Dr. Homi Bhabha Road, Pashan, Pune, India*
[b]*Hefei National Laboratory for Physical Sciences at the Microscale, University of Science and Technology of China, Hefei, Anhui 230026, People's Republic of China*
[c]*UGC-DAE Consortium for Scientific Research, University Campus, Khandwa Road, Indore, India*
[d]*Centre for Energy Science, Indian Institute of Science Education and Research, Pune*

*Corresponding author:smita.chaturvedi24@gmail.com,smita.chaturvedi@iiserpune.ac.in, s.kulkarni@iiserpune.ac.in*

## S1. X-ray Diffraction

**Table S1.** Derived structural parameters

| | $a$ (Å) | $b$ (Å) | $c$ (Å) | $V$ (Å$^3$) | Fe-O1$^{(v)}$ (Å) | Fe-O2$^{(ix)}$ (Å) | Fe-O2 (Å) | Orthorhombic Strain ($s$) | <Fe-O-Fe> | <$\Phi$> |
|---|---|---|---|---|---|---|---|---|---|---|
| SFO-1 | 5.590 | 7.711 | 5.401 | 232.807 | 2.023 | 2.025 | 2.022 | 0.0341 | 146.51 | 16.74 |
| SFO-2 | 5.597 | 7.712 | 5.402 | 233.172 | 1.992 | 2.021 | 2.034 | 0.0353 | 148.39 | 15.80 |
| Single crystal[1] | 5.600 | 7.706 | 5.400 | 233.010 | 2.001 | 2.012 | 2.028 | 0.0365 | 148.65 | 15.67 |
| ANN[2] | 5.569 | 7.694 | 5.378 | 230.435 | 2.000 | 2.015 | 2.003 | - | - | 15.93 |

**\*** *Data compared with single crystal data and theoretical prediction using ANN.*

Figure S1 illustrates Reitveld refined X-ray diffraction patterns for samples SFO-1 and SFO-2.

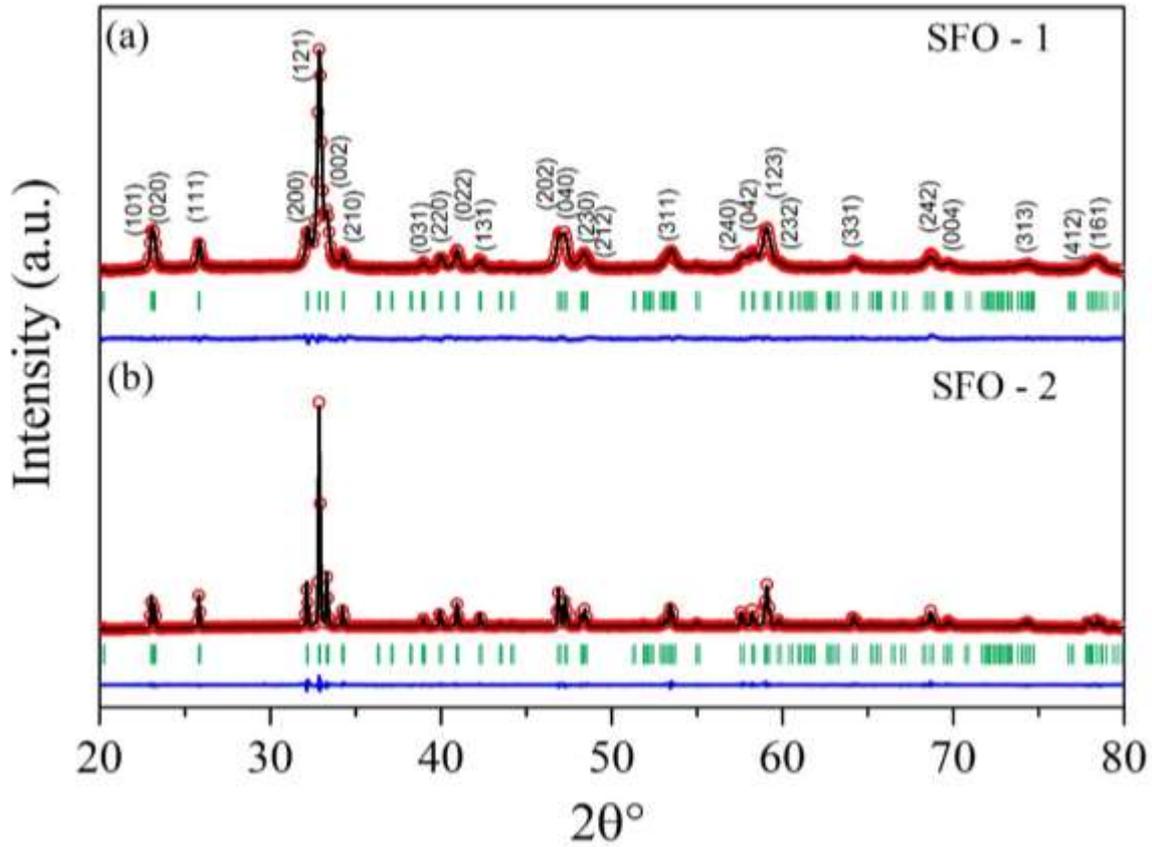

**Fig. S1.** Rietveld-refined X-ray diffraction patterns for samples (c) SFO-1 and (d) SFO-2.

## S2.  Dielectric Measurements

Figure S2 shows the dielectric constant vs. temperature across $T_{SR}$ for both the samples studied. In the ~55 nm particles (Figure S3a), we zoom-in on the earlier reported[3] slope-break in permittivity $\varepsilon'_\omega(T)$ at $T_{SR}$ ~480 K. On the other hand, in the ~500 nm SFO-2's raw data (Figure S3b), larger permittivity), no distinct anomaly is discerned near $T_{SR}$.

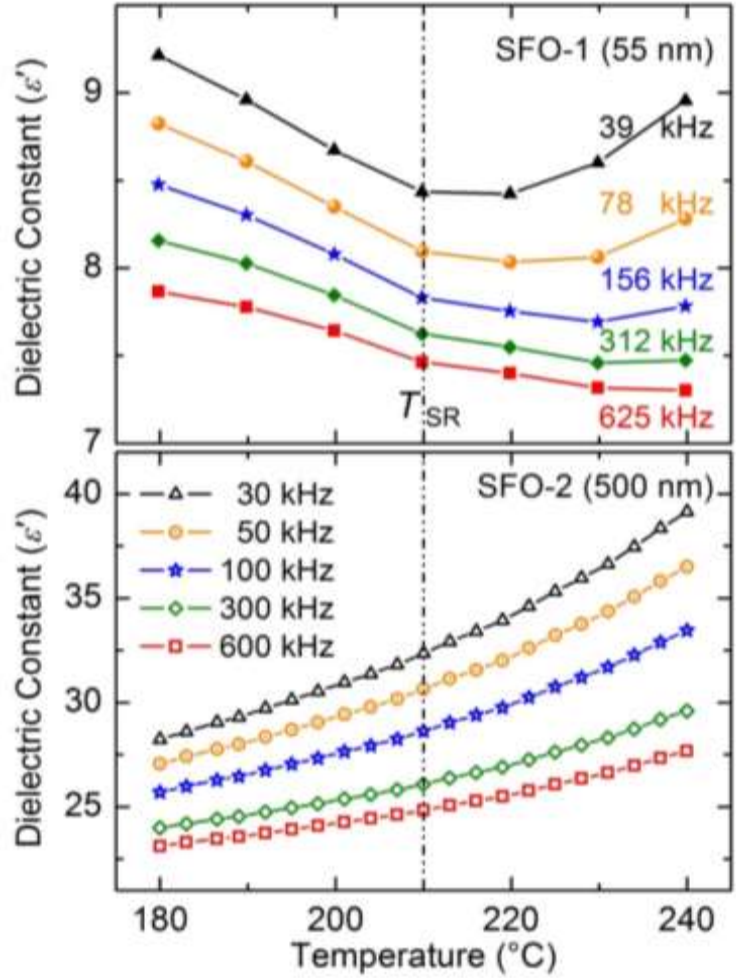

**Fig. S2**. The dielectric constant vs. temperature measured in the SFO samples. The data reveals clear slope-break anomaly at $T_{SR}$ in the SFO-1 (top panel) and its smeared-off/hindered nature in SFO-2 (bottom panel).

### S3. Thermal Characterization

The simple model invoked to reconcile the counter-intuitive size-dependent prominence observed in magneto-electric signatures of the AFM-FE ordering, is apparently corroborated by the heat-capacity thermographs shown in Supplementary Figure S3, for the nano- and micro- grainsized samples. Surprisingly, here as per the usual expectation, the signature of the AFM-FE transition is rather smeared out for SFO-1 and of the AFM sharply-peaked for SFO-2, as consistent with the magnetization-signal-derivative $d\chi/dT$ in Figs. 3a and 3b insets. The circumstance of single (SFO-1) and multiple (SFO-2) FE-domains hosted per grain is nonetheless consistent with this observation. From the standpoint of thermal energy content (enthalpy), SFO-2 has built-in additional-- intra-grain degrees of freedom, with the randomized (incipient) FE-domains within. Therefore, SFO-2 grains comprising multi FE-domains are endowed with extra thermal mass vis-à-vis SFO-1 grains, hosting individual FE-domains. The higher enthalpy of SFO-2 shows up in its more prominent heat capacity peak, upon the AFM ordering. Agreeably, due to the higher bulk-content, the magnetic contribution to the $C_p$-anomaly is also relatively sharper in SFO-2; though this may partially account for the observed thermographs' contrast.

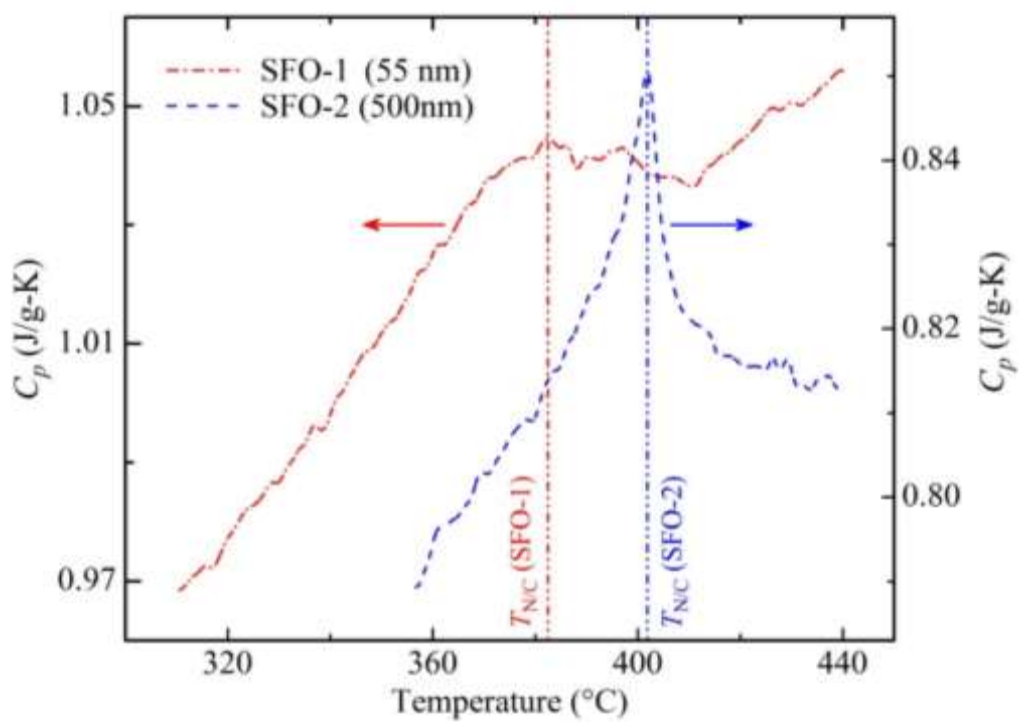

**Fig. S3**. Specific heat thermograms of both the samples. Peaks in SFO-1 (~55 nm, left *y*-axis) and SFO-2 (~500 nm, right *y*-axis) signify the AFM transition. Contrasting nature of the anomalies observed here is consistent with the reckoned assertion of single/multiple FE domains, comprising the grains of the two samples respectively, the circumstance invoked to explain the anomalously contrasted features in Fig.7. The discrepancy between the marked temperatures observed in these thermograms and those marked in the dielectric permittivity (Fig.7) is presently unclear, and requires further investigation.